\documentclass[twocolumn]{revtex4}
\usepackage{amsmath,amssymb,graphicx,bbm}

\newcommand{\tmop}[1]{\ensuremath{\operatorname{#1}}}

\begin{document}

\title{The Spin-Exchange Dynamical Structure Factor of the $S=1/2$ Heisenberg Chain}

\author{Antoine Klauser${}^{1,2}$, Jorn Mossel${}^2$, Jean-S\'ebastien Caux${}^2$ and Jeroen van den Brink${}^3$}

\affiliation{${}^1$Instituut-Lorentz, Universiteit Leiden, P. O. Box 9506, 2300 RA Leiden, The Netherlands}
\affiliation{${}^2$Institute for Theoretical Physics, Universiteit van  
Amsterdam, 1090 GL Amsterdam, The Netherlands}
\affiliation{${}^3$Institute for Theoretical Solid State Physics, IFW Dresden, 01171 Dresden, Germany}
\date{\today}

\begin{abstract}
We determine the spin-exchange dynamical structure factor of the Heisenberg spin chain, as is measured by indirect Resonant Inelastic X-ray Scattering (RIXS). We find that two-spin RIXS excitations nearly entirely fractionalize into {\it two spinon} states. These share the same continuum lower bound as single-spin neutron scattering excitations, even if the relevant final states belong to orthogonal symmetry sectors. The RIXS spectral weight is mainly carried by higher-energy excitations, and is beyond the reach of the low-energy effective theories of Luttinger liquid type. 
\end{abstract}

\maketitle

A remarkable feature of quantum systems in one dimension (1D) is that quantum criticality is the norm rather than the exception~\cite{GiamarchiBOOK}. 1D and quasi-1D systems as diverse as carbon nanotubes, stripes in cuprate high-temperature superconductors, confined ultracold atomic gases and quantum spin chains all provide realizations of critical quantum liquids. Possibly the most studied prototypical 1D quantum critical system is the Heisenberg $S=1/2$ antiferromagnetic chain~\cite{1928_Heisenberg_ZP_49}. Its basic excitations are spinons, fractionalized spin excitations that emerge in the critical state~\cite{1981_Faddeev_PLA_85}.
Until recently, the only technique available to investigate these fractional spin excitations was inelastic neutron scattering (INS). In neutron scattering integer spin-flip excitations carrying S=1 are created, so that the INS response involves the pairwise creation of spinons. The INS amplitude is determined by the single-spin dynamical structure factor (DSF) and a precise matching between the observed INS intensity and the dynamical structure factor calculated from theory has recently been achieved~\cite{2007_Kohno_NATPHYS_3,2009_Walters_NATPHYS_5,2009_Thielemann_PRL_102}. 

A few years ago, x-ray photon scattering emerged as a new tool to measure dispersive magnetic excitations~\cite{Ament2011}. When the incident photon energy is resonant with an absorption edge  of the material, yielding Resonant Inelastic X-ray Scattering (RIXS), the correlations between two neighboring spins factor into the magnetic x-ray scattering processes. The corresponding scattering amplitude is given by the momentum dependent {\it two}-spin dynamical structure factor, also referred to as the {\it spin-exchange} DSF. When for instance in an antiferromagnetic copper-oxide compound the incident energy is tuned to the copper K-edge -- so-called {\it indirect RIXS} -- exclusively this type of magnetic scattering occurs~\cite{Ament2011,2007_van_den_Brink_EPL_80, 2008_Forte_PRB_77,Hill2008,2010_Ellis_PRB_81}. The recent experimental advances in inelastic x-ray scattering allow to probe one of the most elementary, fractionalized states of critical matter in an entirely new manner, highlighting more elaborate spin-spin correlations.

It is the purpose of this Letter to provide a nonperturbative calculation of the spin-exchange RIXS scattering amplitude for the Heisenberg $S=1/2$ antiferromagnetic chain 
\begin{eqnarray}
H = J \sum_i \left( {\bf S}_i \cdot {\bf S}_{i+1} - 1/4 \right).
\label{eq:Hamiltonian}
\end{eqnarray}
Our results aim on the one hand to challenge, guide and inspire further experimental magnetic RIXS efforts; 
on the other hand the response function that we consider here is of fundamental significance and, as explicitly shown later on, beyond the reach of traditional low-energy based theories (such as e.g. Luttinger liquid theory) because it sits at energies of order of the exchange $J$. Our problem and approach thus represent an uncommon example of direct contact between experiment and theory beyond universality.

The model's integrability~\cite{1931_Bethe_ZP_71,GaudinBOOK,KorepinBOOK,TakahashiBOOK} is well-known to permit nonperturbative calculations of equilibrium properties, but it has recently also become possible to accurately compute certain dynamical properties for both finite~\cite{2003_Biegel_JPA_36,2004_Sato_JPSJ_73,2005_Caux_PRL_95,2005_Caux_JSTAT_P09003,2009_Kohno_PRL_102} and infinite \cite{1995_Jimbo_Book,1997_Karbach_PRB_55,2006_Caux_JSTAT_P12013,2008_Caux_JSTAT_P08006} systems. Until now these predictions have been limited to the single-spin DSF measured by INS. For experiments other than INS, governed by different correlators, at present no results are available. Here we fill this gap for RIXS, using the exact Algebraic Bethe Ansatz to calculate the spin-exchange DSF. 

\begin{figure}
\includegraphics[width=.85\columnwidth]{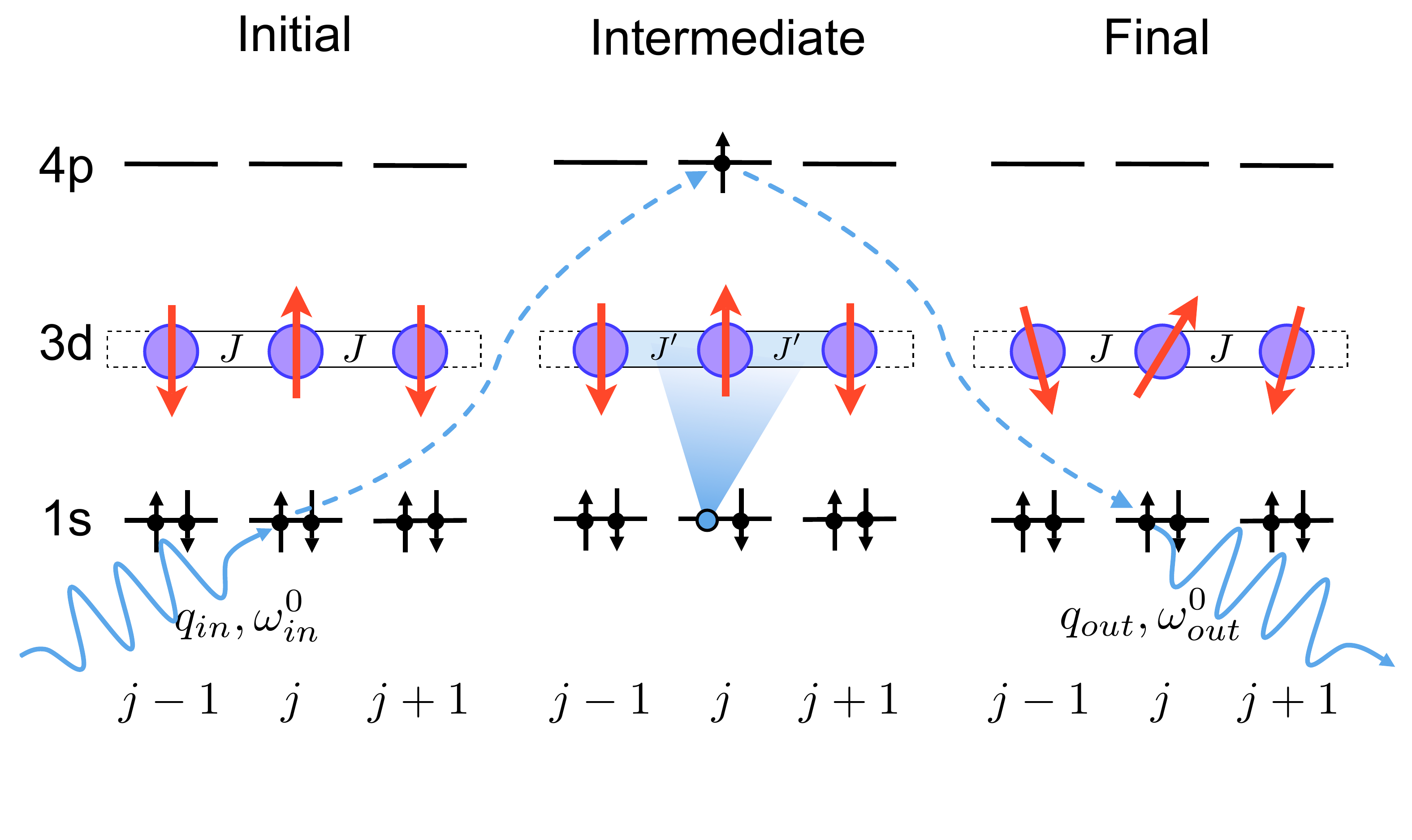}
\caption{Mechanism by which double-spin flip transitions are created in the indirect magnetic RIXS process.}
\label{fig:RIXS}
\end{figure}

{\it Magnetic RIXS cross section --}
Before calculating its response function, we briefly describe the magnetic RIXS process. RIXS is a photon-in photon-out scattering technique in which the energy of the incident x-ray photons is tuned to an atomic absorption edge of the material that is studied. X-rays can for instance be tuned to the $K$-edge of a transition metal ion ($1-10 keV$), for example copper, in the sample. In this case a magnetic scattering process as sketched in Fig.~\ref{fig:RIXS} can occur  \cite{2007_van_den_Brink_EPL_80,2008_Forte_PRB_77}. An x-ray incoming on site $j$ produces a $1s-4p$ electronic transition and creates a core-hole on the $j$th copper ion of the chain. In the intermediate state, the presence of the core-hole modifies the $3d$ on-site energy levels through the Coulomb interaction. In the Mott-insulating limit, this perturbation modifies the spin-exchange process with the two neighboring $3d$ electrons, and thus locally modifies the actual superexchange coupling $J$ between the spins. Following the notation of \cite{2007_van_den_Brink_EPL_80,2008_Forte_PRB_77} we denote the perturbed coupling $J^c = (1+\eta)J$. In the final state, the core-hole is filled again by the $4p$ electron, but the spin chain is left behind in an excited state. The cross section for this scattering process is given by the Kramers-Heisenberg relation.  As a function of energy loss $\omega=\omega^0_{in}-\omega^0_{out}$ and momentum transfer along the spin chain $q = q_{in}-q_{out}$ the scattering intensity is
$
I=\sum_f \left| A_{f i} \right|^2 \delta (w - E_f +E_i),
$
where the scattering amplitude is
$
A_{f i} = \omega_{\tmop{res}} \sum_n \frac{\left\langle f \right| \hat{D} 
   \left| n \right\rangle \left\langle n \right| \hat{D} \left| i
   \right\rangle}{\omega_{\tmop{in}} - E_n - i \Gamma}. 
$
Here $| i \rangle$, $| n \rangle$, $| f \rangle$  are the initial, intermediate and final states with respective energies $E_i$, $E_n$, $E_f$ and $\hat{D}$ is the dipole operator that creates or annihilates photo-excitations. Because the $1s$ core-hole is highly energetic, it quickly decays, leading to an energy broadening $\Gamma$ of the intermediate state -- its inverse lifetime.
For $\Gamma\gg E_n$, $A_{f i}$ can be expanded in a power series which re-summed to leading-order ultimately provides the x-ray  scattering cross section 
$
I \propto
    \left | \frac{\omega_{\tmop{res}}}{ \Gamma} \frac{ \eta J} {i
   \Gamma + \omega}\right|^2 S^{exch}(q,\omega).
$
It incorporates the spin-exchange dynamical structure factor
\begin{eqnarray}
S^{exch}(q,\omega) = 2 \pi \sum_{\alpha} \left|\left\langle 0 \right| 
X_{q}
\left| \alpha \right \rangle  \right|^2\delta(\omega-\omega_\alpha),
\label{eq:DSF}
\end{eqnarray}
where the ground state is $| 0 \rangle$, the excited states $| \alpha \rangle$, excitation energies $\omega_{\alpha } = E_\alpha-E_{GS}$ and the spin-exchange operator is
$X_{q} \equiv \frac{1}{\sqrt{N}} \sum_j e^{i q j}(\mathbf{S}_{j-1}\cdot \mathbf{S}_{j}+\mathbf{S}_{j}\cdot \mathbf{S}_{j+1})$.
Motivated by the successful correspondence between predictions and experiments for the long-range ordered 2D Heisenberg antiferromagnet \cite{2008_Forte_PRB_77,2010_Ellis_PRB_81,Hill2008,2009_Braicovich_PRL_102}, we set out to compute this structure factor in the quantum critical 1D case, applicable for instance to $\mbox{Sr}_2\mbox{CuO}_3$. 

\begin{figure}
\includegraphics[width=\columnwidth]{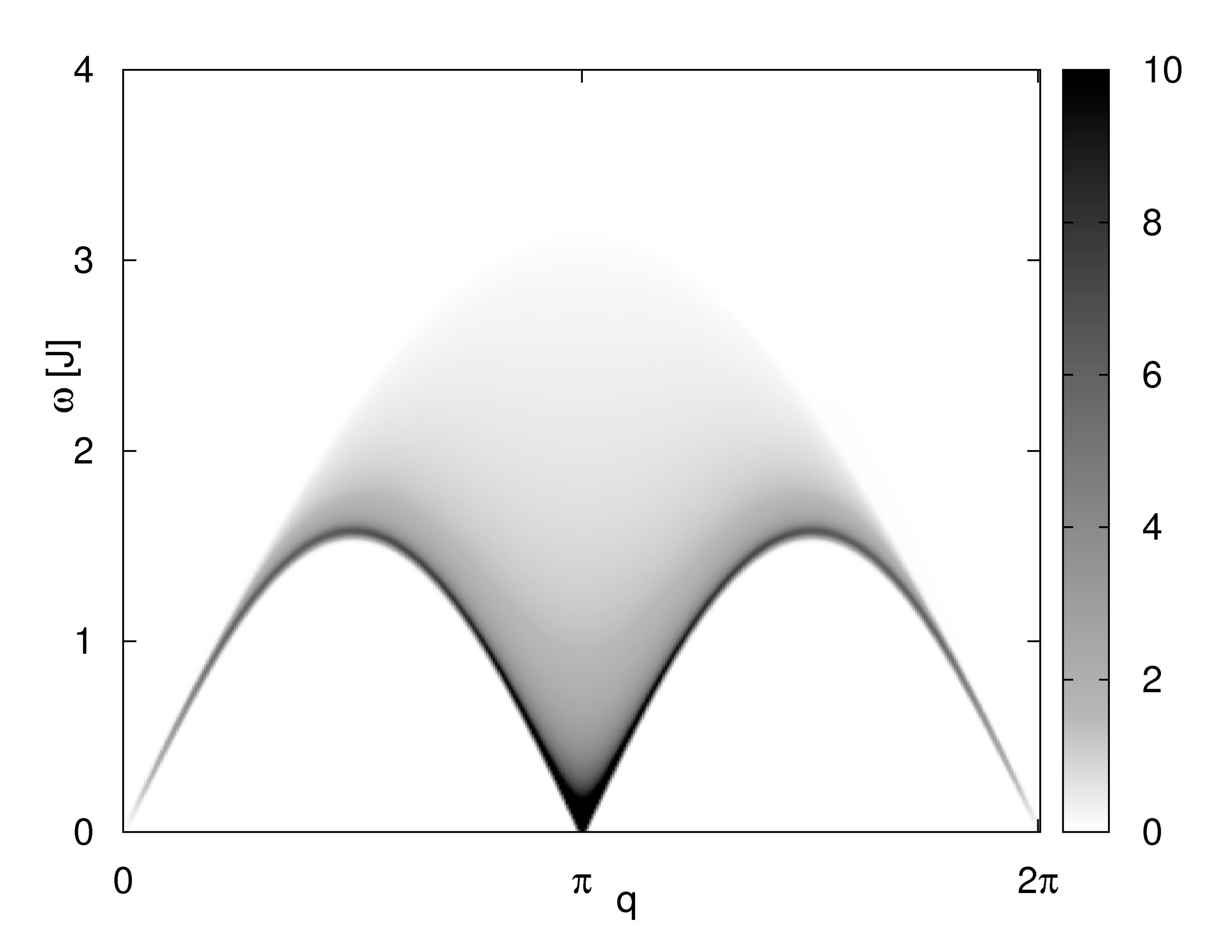}
\caption{Single spin dynamical structure factor $S^{single}_{zz}(q,\omega)$ of the Heisenberg chain. The DSF is dominated by the $2$-spinon spectrum but also 4-spinon and higher excited states contribute ($6 \% $). The computation is for $N=400$ sites. }
\label{fig:Sz}
\end{figure}

{\it Computing the spin-exchange DSF --}
So far exact calculations for the spin chain have been restricted to the single-spin DSF, as measured in neutron scattering: 
\begin{equation}
S^{single}_{a\overline{a}}(q,\omega)= 2\pi \sum_\alpha | \langle 0 | S^a_q |  \alpha \rangle |^2 \delta(\omega-\omega_\alpha),
\label{eq:INSDSF}
\end{equation} 
where $ \langle 0 | S^a_q |  \alpha \rangle$ is the form factor (FF) 
of the Fourier transformed spin operators $S^a_q=(1/ \sqrt{N})\sum_{j=1}^Ne^{-\mathrm{i} qj}S^a_j$, $a = z,+,-$ between  the ground state and eigenstates of excitation energy $\omega_{\alpha}$ (see Fig.~\ref{fig:Sz}).
The strategy which we adopt for the calculation of the RIXS response function $S^{exch}$ (Eq.~\ref{eq:DSF}) is that of the ABACUS method \cite{2009_Caux_JMP_50}. Within this approach eigenstates are explicitly obtained from Bethe Ansatz, matrix elements from the Algebraic Bethe Ansatz, and the trace over intermediate states is performed numerically using an optimized search through the Hilbert space.  In what follows, we briefly describe each of these three steps for the specific case of the RIXS response function (Eq.~\ref{eq:DSF}).

The Hilbert space can be divided into subspaces of fixed magnetization characterized by the number of downturned spins $M$. Eigenstates of Eq.~\ref{eq:Hamiltonian} for an (even) $N$ sites periodic spin chain are completely determined for $M \leq \frac{N}{2}$ by a set of $M$ rapidities $\{ \lambda_1, \ldots, \lambda_M \} $ which for the isotropic chain solve the Bethe equations
\begin{equation}
 \arctan \left(2 \lambda_i\right) =\frac{\pi}{N}I_i + \frac{1}{N} \sum_{k=1}^{M}\arctan \left(\lambda_i-\lambda_k\right) 
\end{equation} with  ${I_1,\ldots, I_M}$ a set of integers for odd $M$ and half-odd integers for even $M$. Each set of quantum numbers specifies a set of rapidities. The ground state is defined by $\{ I^0_k=k -\frac{M+1}{2} \},\> k=1,\ldots,M$. The energy and the momentum of an eigenstate are given by
\begin{eqnarray}
E =-J \sum_{k=1}^{M}\frac{1/2}{1/4+\lambda_k^2}; \ \ 
 P =\pi M -\frac{2 \pi}{N}\sum_{k=1}^{M} I_k  \left( {\rm mod} \ 2\pi\right). \nonumber
\end{eqnarray}

\begin{figure}
\includegraphics[width=\columnwidth]{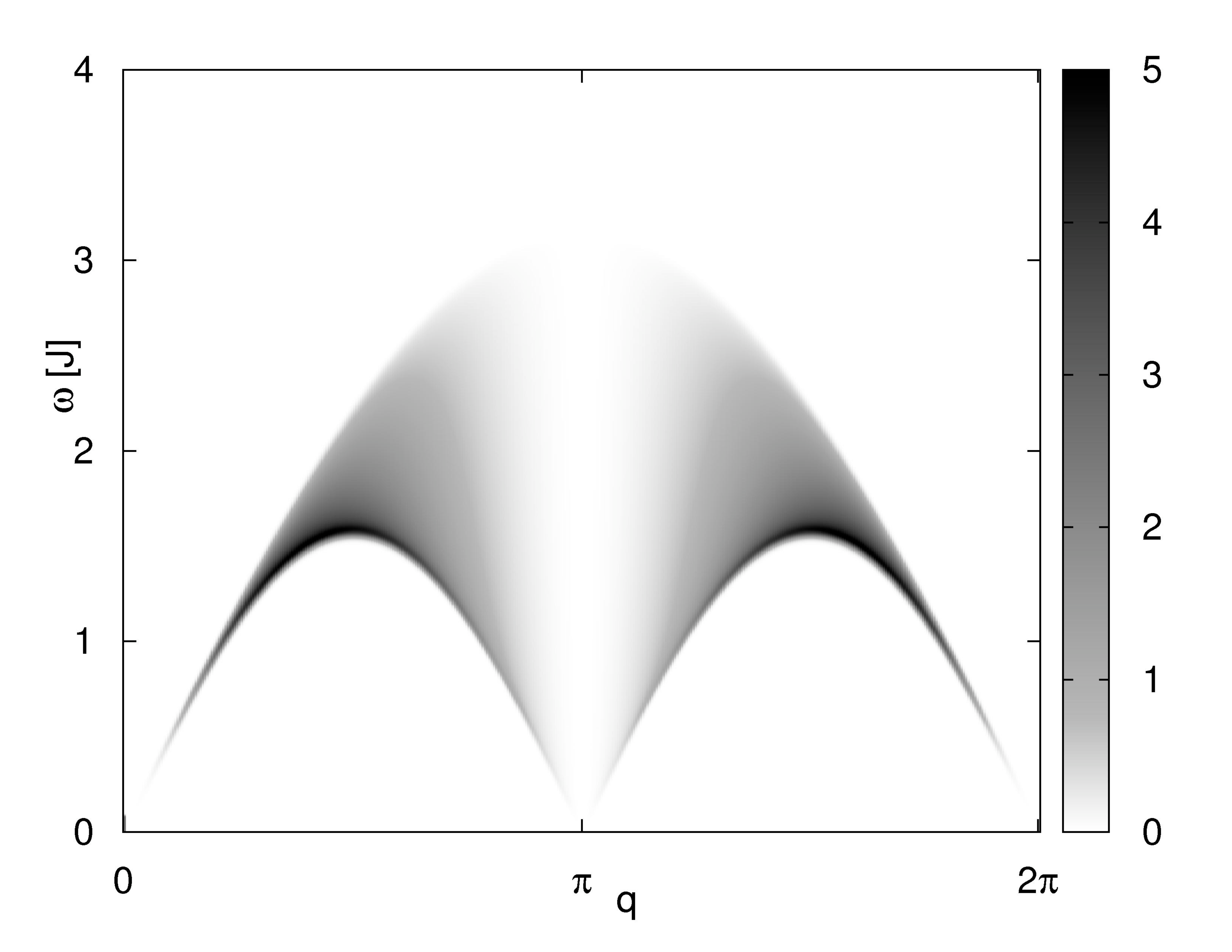}
\caption{Spin-exchange dynamical structure factor $S^{exch}$ of the isotropic Heisenberg chain. The sharp on-set of the spinon continuum is very similar to the single-spin DSF $S^{single}_{zz}$. The intensity around $q= \pi$ is suppressed, thereby enhancing high-energy response around $q= \frac{\pi}{2},\frac{3 \pi}{2}$. The computation is for $N=400$ sites.}
\label{fig:SzSz}
\end{figure}

Turning now to the matrix elements of the exchange operators $X_q$, we can exploit the spin isotropy of the system to express the full spin-exchange DSF matrix element in Eq.~\ref{eq:DSF} as a function of the $S^z_{i}S^z_{i+1}$ FFs only.  By globally rotating the $ \left\langle 0 \right|S^x_{j}S^x_{j+1} \left| \alpha \right \rangle$ and $ \left\langle 0 \right|S^y_{j}S^y_{j+1} \left| \alpha \right \rangle$ FFs appearing in Eq.~\ref{eq:DSF} about the $y$ and $x$ spin axes, respectively, and using the fact that the ground state is a global $su(2)$ singlet, one can show that only singlet excited states contribute to Eq.~\ref{eq:DSF}. This conclusion is reached by first noticing that the $S^z_{i}S^z_{i+1}$ operator creates excited state only in the $S_{tot}=0,1,2$ sectors. Secondly, the rotation of states belonging to these sectors and with zero magnetization ($S^z_{tot}=0$) gives
\begin{eqnarray}
&&e^{i\phi\sum_j S_j^x}| \alpha, S^{z}_{tot}=0,S_{tot}=0  \rangle = | \alpha, 0,0  \rangle \nonumber\\
&&e^{i\phi \sum_j S_j^x}| \alpha, 0, 1  \rangle = \cos (\phi)| \alpha, 0,1  \rangle + \ldots \nonumber\\
&&e^{i\phi \sum_j S_j^x}| \alpha, 0, 2  \rangle = \frac{1 + 3\cos (2\phi)}{4}| \alpha, 0,2  \rangle + \ldots \nonumber
\label{eq:rotation}
\end{eqnarray}
(in which $\dots$ represent $S^z_{tot} \neq 0$ states of which the $S^z_j S^z_{j+1}$ matrix elements with the ground state vanish) with similar results for a rotation around the $y$ axis. With rotations of respectively $\pm \pi/2$ to transform either $S_j^y$ or $S_j^x$ into $S_j^z$, the total contribution of eigenstates with $S_{tot}=1,2$ vanishes, allowing one to rewrite the spin-exchange DSF as
\begin{eqnarray}
&&S^{exch}(q,\omega) = \cos^2(q/2)\frac{72 \pi}{N} \times \nonumber \\
&&\sum_{\alpha \in S_{tot}=0}\sum_j \left| e^{i q j  }\left\langle 0 \right| S^z_{j} S^z_{j+1} \left| \alpha \right \rangle  \right|^2\delta(\omega-\omega_\alpha).
\label{eq:DSFzz}
\end{eqnarray}
In order to calculate the matrix elements $\left\langle 0 \right| S^z_{j} S^z_{j+1} \left| \alpha \right \rangle$, we make use of Algebraic Bethe Ansatz methods \cite{KorepinBOOK,1989_Slavnov_TMP_79,1990_Slavnov_TMP_82,1999_Kitanine_NPB_554,2000_Kitanine_NPB_567} to represent these as explicit functions (taking the form of matrix determinants) of the rapidities involved in the left and right eigenstates. State norms are also given by an explicit determinant \cite{GaudinBOOK,1981_Gaudin_PRD_23,1982_Korepin_CMP_86}. For brevity, explicit expressions are not presented here \cite{Klauser_TBP}.
\begin{figure}
\includegraphics[width=.85\columnwidth]{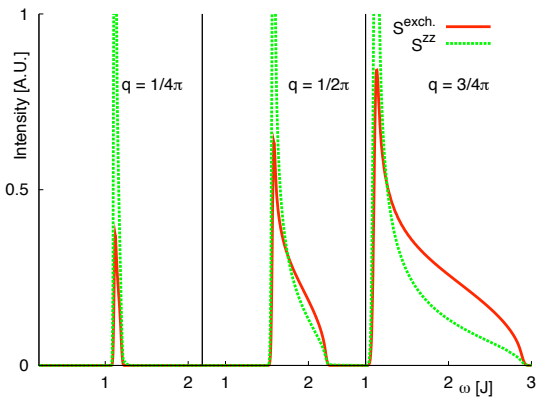}
\caption{Fixed momentum profiles of the spin-exchange dynamical structure factor $S^{exch}(q,\omega)/(\cos^2 (q/2))$ (plain) and $S^{single}_{zz}(q,\omega)$ (dashed), each normalized to its own sum rule. Data are plotted as function of the energy loss $\omega$, at $q=\pi/4$, $\pi/2$, $3\pi/4$.}
\label{fig:Profiles}
\end{figure}

The third step, summation over intermediate states, can now be performed to obtain quantitative results for the DSF in Eq.~\ref{eq:DSF}.  This is done using the ABACUS algorithm \cite{2009_Caux_JMP_50} which sums intermediate state contributions in a close to optimal order. 
By solving the Bethe equations, the eigenstate rapidities give the energy,  momentum and the FF value through the determinant expression. 
Large families of excited states are summed over until satisfactory saturations of sum rules are achieved. More precisely, first, we compare the computed integrated intensity with the analytical result in the infinite chain limit: $\int\frac{d\omega}{2\pi}\frac{1}{N}\sum_{q}S^{exch}(q,\omega) = \frac{1}{4}-\ln(2)+\frac{9}{8}\zeta(3)$. A second check is the first frequency moment sum rule $\int\frac{d\omega}{2\pi}\omega S^{exch}(q,\omega)= 6 \sin^2(q) \left\{ (x_1 - x_2) \left(1- 4\cos^2 (q/2) \right) + \frac{3 \zeta(3)-4\ln(2)}{8} \right \}$ (valid for the isotropic case, in which $x_i \equiv \langle S_j S_{j+i} \rangle$; we have made use of the results of \cite{2003_Sakai_PRE_67}). To obtain smooth curves in frequency $\omega$, the delta function in Eq.~\ref{eq:DSF} is broadened to a scale commensurate with the energy level spacing. The results for the RIXS scattering amplitude are presented in Fig.~\ref{fig:SzSz} (for all momenta in the full Brillouin zone) and \ref{fig:Profiles} (as a function of energy at fixed momentum). 

As the RIXS response involves two spin excitations, to each of which a $S=1$ is associated, one might naively expect it to be dominated by {\it four} spinon intermediate states. However, it turns out that the RIXS excitation fractionalizes almost completely into {\it two} spinons. The state contributions to the sum-rule shows that the two-spinon states which cover all but $ 10^{-4} \% $ of the RIXS signal for a finite size chain with $N=400$. This ratio is even more drastic than for the INS results where the excitation of four and more spinons is responsible for $ 6\% $ ($N=400$) of the signal. Interestingly, the magnetic RIXS response is at first glance rather similar to the neutron scattering one, sharing the same lower bound of the spinon continuum, even if the RIXS and INS final states belong to different symmetry sectors and are orthogonal. The fact that the spin-exchange DSF exists within a continuum is a direct and explicit consequence of the fractionalization of spin excitations in the quantum critical spin chain. Even if the excitation continuum probed by RIXS exactly coincides with the one probed by INS, the spectral weight has a markedly different distribution. This crucial difference is partly caused by the static factor $\cos^2(q/2)$ which originates from the modification of two neighboring exchange couplings in the x-ray scattering process. Excitations are thus associated to a typical length $2a$ (with $a$ the lattice spacing) or equivalently predominantly carry $\pm\frac{\pi}{2}$ (mod $2\pi$) momentum. The figures clearly illustrate this fact: for INS (Fig.~\ref{fig:Sz}) the signal is maximal at the antiferromagnetic wavevector $q = \pi$ at $\omega$ close to zero, whereas the RIXS amplitude vanishes there. Rather it is concentrated above the continuum threshold at $q = \pi/2$, characterized by a vanishing group velocity. The fixed-momentum profiles in Fig.~\ref{fig:Profiles} show further differences between RIXS and INS apart from the $\cos^2(q/2)$ factor. While the weight of both responses predominantly sits between the top and bottom of the two-spinon continuum, the weight distribution between these two are clearly distinct, the RIXS response having a much broader shoulder at higher $\omega$ than the INS response.  This quantitative difference is sufficiently large to be observable experimentally.

The asymptotic behavior of correlations along the spin chain are well described by low energy effective theories \cite{GiamarchiBOOK}, which provide a detailed understanding of the antiferromagnetic singularity at momentum $\pi$ and zero energy in the INS response. The exponent of the singularity at the lower bound of the spectrum is given by the asymptotic decay of the two-spin correlation. Straightforward application of  the known results for the four-spin correlator from Luttinger liquid theory \cite{1989_giamarchi_prb_39} to the RIXS intensity fails as the signal vanishes at all low energies. The development nonlinear extensions to Luttinger liquid theory \cite{2009_Imambekov_SCIENCE_323} or alternately the quantum group approach \cite{1995_Jimbo_Book,1997_Karbach_PRB_55,2006_Caux_JSTAT_P12013,2008_Caux_JSTAT_P08006} might bring this within reach in the future. 

We have, in conclusion, explicitly calculated the relevant magnetic response function for indirect RIXS on the Heisenberg chain, which involves the $\mathbf{S}_i\mathbf{S}_{i+1}$ spin-exchange operators. This operator predominantly probes high-energy magnetic excitations and the integrability-based method which we have employed here is then the only one capable of describing RIXS. Alternate approaches based on approximate low-energy theories cannot reach energies of order of the exchange $J$. We have demonstrated in more general terms that via the Algebraic Bethe Ansatz formalism a new family of correlators of the type $S^a_iS^b_{i+1}$ with $a,b = z,+,-$ is accessible for computation. 
The challenges of future work are to extend the present exact calculation of the spin-exchange dynamical structure factor to anisotropic spin chains, including also finite magnetic fields.

\end{document}